\documentstyle[12pt]{article}
\input psfig.tex
\input tex.def

\def\etal{{{\it et al. }}}
\newcommand{\farcs}{\hbox{$.\!\!^{\prime\prime}$}}
\def\title#1{\relax\vspace*{2cm}{\large{\bf #1}}\par\vspace*{13.5pt}}
\def\author#1{{#1}\par\vspace*{13.5pt}}
\def\affil#1{{\it #1}\par}
\def\abstract{\vspace*{27pt}ABSTRACT\par\relax}
\def\section#1{\par{#1}\par}
\def\subsection#1{\par\underline{#1}\par}
\def\subsubsection#1{\par\underline{#1.}\ \ }

\def\acknow{\par ACKNOWLEDGMENTS\par}
\newenvironment{references}{\section{REFERENCES}\vspace*{.5cm}%
\parindent=0pt\frenchspacing%
\parskip=1pt plus 1pt minus 1pt%
\interlinepenalty=1000\tolerance=400%
\pretolerance=10000\hyphenpenalty=10000%
\everypar={\hangindent=1.6pc}
}{}
 
\begin{document}

\title{LINERs AS LOW-LUMINOSITY ACTIVE GALACTIC NUCLEI\footnote{Invited review
to appear in {\it The 32$^{nd}$ COSPAR Meeting, The AGN-Galaxy Connection}
(Advances in Space Research).}}

\author{Luis C. Ho$^{\dag,\ddag}$}
\affil{$^{\dag}$Harvard-Smithsonian Center for Astrophysics
60 Garden St., Cambridge, MA 02138, USA}
\affil{$^{\ddag}$Carnegie Observatories, 813 Santa Barbara St.
Pasadena, CA 91101-1292, USA}

\abstract
Many nearby galaxies contain optical signatures of nuclear activity in the 
form of LINER nuclei.  LINERs may be the weakest and most common manifestation
of the quasar phenomenon.  The physical origin of this class of objects, 
however, has been ambiguous.  I draw upon a number of recent observations 
to argue that a significant fraction of LINERs are low-luminosity 
active galactic nuclei.

\section{AGN CENSUS IN NEARBY GALAXIES}

The local space density of active galactic nuclei (AGNs) has bearing on a 
number of issues in extragalactic astronomy, including the 
fraction of galaxies hosting massive black holes, the cosmological evolution 
of quasars, and the contribution of AGNs to the cosmic X-ray background.  
It is therefore of fundamental importance to establish the extent and nature 
of nuclear activity in nearby galaxies.  This contribution summarizes recent 
efforts to survey nearby galactic nuclei, discusses complications regarding the 
interpretation of the results, and presents a variety of fresh observational 
perspectives that help toward reaching a coherent understanding of nuclear 
activity in nearby galaxies.

Optical surveys find that a large fraction of nearby galaxies have nuclei that 
emit weak emission lines with a spectrum unexpected for photoionization by 
normal stars.  Heckman (1980) identified low-ionization nuclear emission-line 
regions (LINERs) as a major constituent of the extragalactic population, 
particularly among early-type galaxies.  The optical spectra of LINERs broadly
resemble those of traditional AGNs such as Seyfert nuclei, except that they 
have characteristically lower ionization levels. These findings were 
strengthened by a number of subsequent studies, as reviewed by Ho (1996).

The latest and most sensitive survey of this kind was completed by Ho \etal
(1997a, and references therein) using the Hale 5~m 
telescope at Palomar Observatory.  Long-slit spectra of exceptional quality 
were taken of the nuclear region of a magnitude-limited ($B_T\,\leq$ 12.5 mag) 
sample of 486 northern ($\delta\,>$ 0\deg) galaxies that constitutes an 
excellent representation of the typical nearby galaxy population.  The spectra 
are of moderate resolution (full-width at half maximum [FWHM] $\sim$ 
100--200 \kms) and cover two regions of the optical window (4230--5110 \AA\ 
and 6210--6860 \AA) containing important diagnostic emission lines.  The 
main results of the Palomar survey are the following.
(1) AGNs are very common in nearby galaxies (Fig. 1).  At least 40\% of all 
galaxies brighter than $B_T$ = 12.5 mag emit AGN-like spectra.  The 
emission-line nuclei are classified as Seyferts, LINERs, or LINER/H~II-region 
composites, and most have very low luminosities compared to traditionally 
studied AGNs.  The luminosities of the H$\alpha$ emission line range from 
10$^{37}$ to 10$^{41}$ ergs s$^{-1}$, with a median value of $\sim$10$^{39}$ 
ergs s$^{-1}$.
(2) The detectability of AGNs depends strongly on the morphological type of 
the galaxy, being most common in early-type systems (E--Sbc).  
The detection rate of AGNs reaches 50\%--75\% in ellipticals, lenticulars, 
and bulge-dominated spirals but drops to \lax 20\% in galaxies classified as 
Sc or later.
(3) LINERs make up the bulk (2/3) of the AGN population and a sizable 
fraction (1/3) of all galaxies.
(4) A significant number of objects show a faint, broad 
(FWHM $\approx$ 1000--4000 km s$^{-1}$) H$\alpha$ emission line that 
 qualitatively resembles emission arising from the conventional broad-line 
region of ``classical'' Seyfert 1 nuclei and QSOs.

\begin{figure}
\vbox{
\vbox{
\hskip 0.1truein
\psfig{file=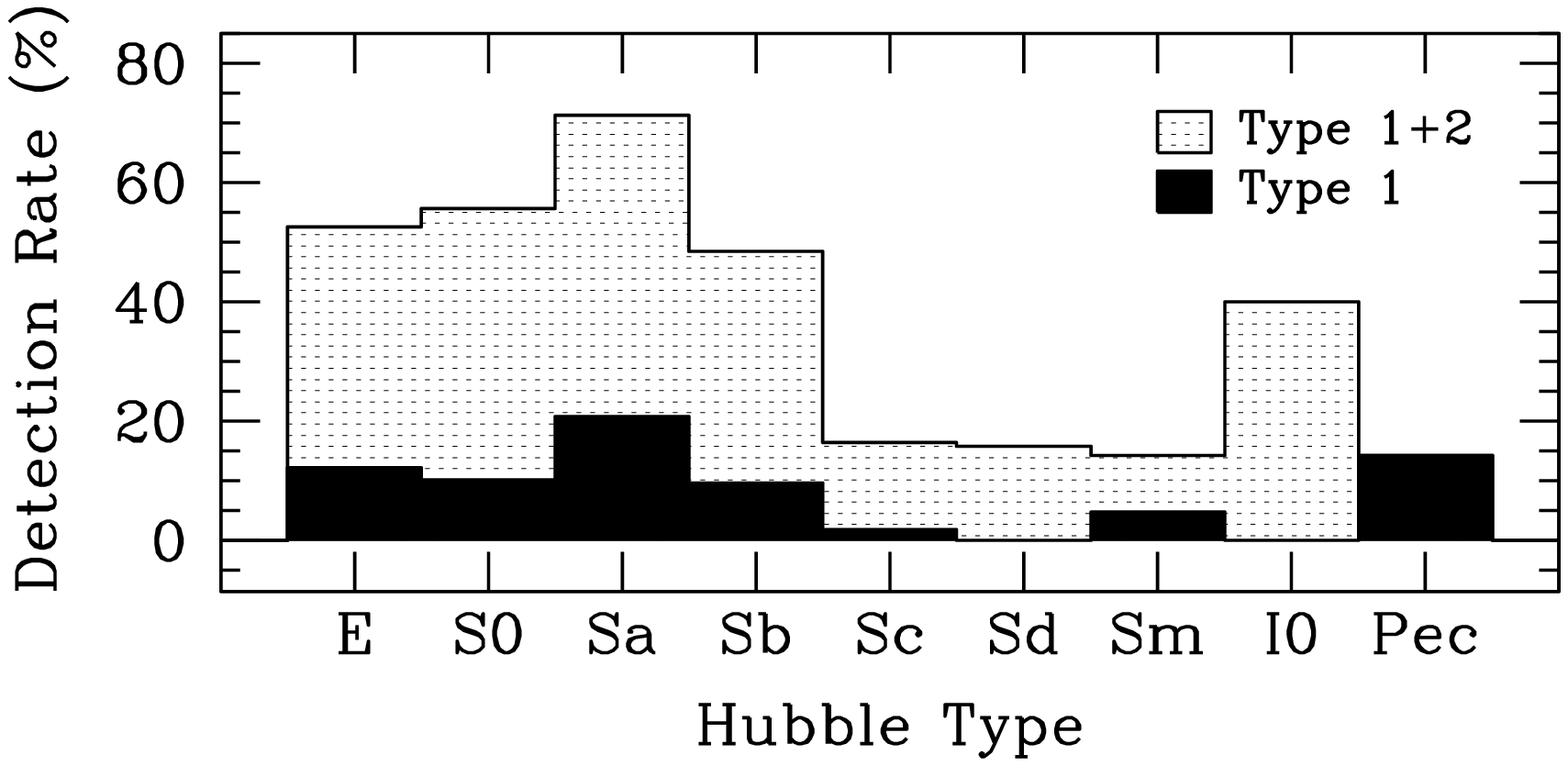,height=1.8truein,width=3.3truein,angle=0}
\hskip 0.1truein
\psfig{file=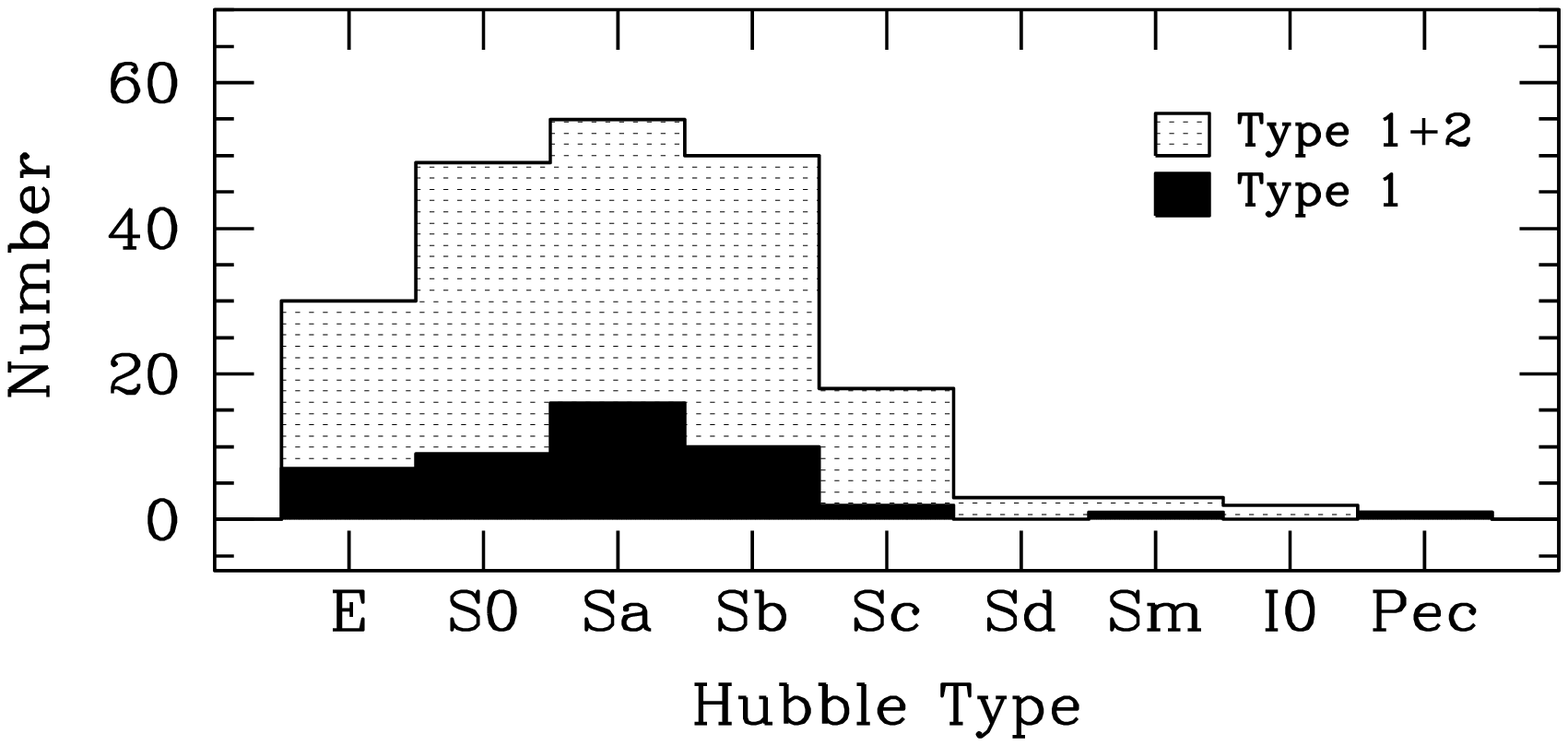,height=1.8truein,width=3.3truein,angle=0}
}
}
\vskip +1.3cm
Fig. 1.  Detection rate ({\it left}) and number distribution ({\it right}) of 
AGNs as a function of Hubble type in the Palomar survey.  ``Type 1'' AGNs 
(those with broad H$\alpha$) are shown separately from the total population 
(types 1 and 2).
\end{figure}

Radio observations provide further support for the prevalence of nuclear 
activity (see the contribution of E. Sadler in these proceedings).  Weak radio 
cores with powers of 10$^{19}$--10$^{21}$ W Hz$^{-1}$ at 5~GHz are found in 
$\sim$50\% of nearby elliptical and S0 galaxies (Sadler \etal 1989; Wrobel and 
Heeschen 1991).  Where information is available, the cores have 
relatively flat spectral indices and nonthermal brightness temperatures 
(Slee \etal 1994), and the optical spectra of most of these sources are 
classified as LINERs (Sadler \etal 1989; Ho 1998a).

\section{RECENT OBSERVATIONAL RESULTS ON LINERs}

\begin{figure}
\hskip 1.2truein
\psfig{file=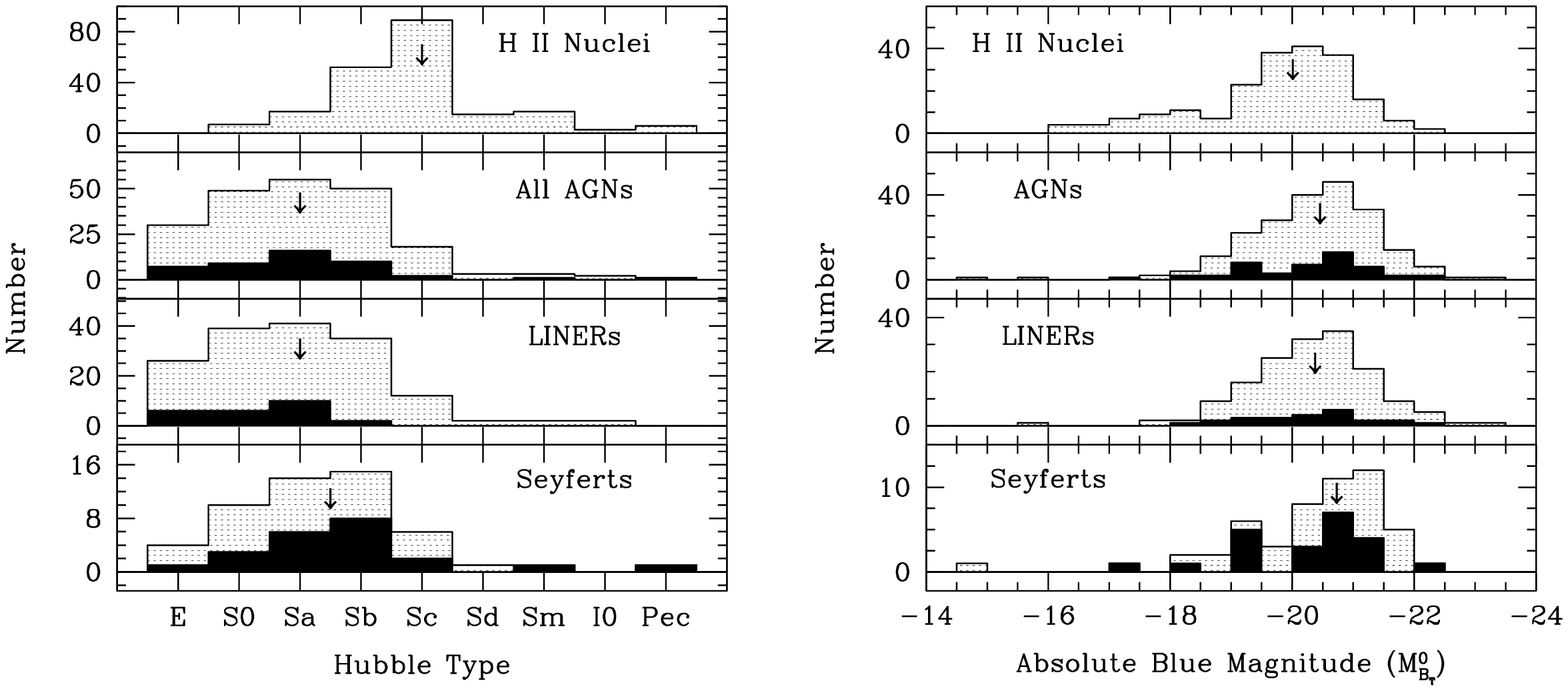,height=3.0truein,angle=90}
\vskip +0.5cm
Fig. 2.  Number distribution of morphological types ({\it left}) and 
total absolute blue magnitudes ({\it right}) for H~II nuclei, all AGNs (LINERs 
+ Seyferts), and LINERs and Seyferts separately.  The median of each 
distribution is marked by an arrow. Adapted from Ho \etal (1997a).
\end{figure}

If LINERs are powered by a nonstellar source, then LINERs clearly would be
the most common type of AGNs known in the nearby regions of the universe.
However, ever since their discovery, the physical origin of LINERs 
has been hotly debated.  The recognition and definition of LINERs is 
based on their spectroscopic properties at optical wavelengths.  In addition 
to the AGN scenario, the optical spectra of LINERs unfortunately can be 
interpreted in several other ways that do not require an exotic energy 
source (e.g., shocks, hot stars; see Ho \etal 1993 and 
Filippenko 1996 for reviews).  As a consequence, it has often been suggested 
that LINERs may be a mixed-bag, heterogeneous collection of objects.
While the nonstellar nature of some well-studied LINERs is incontrovertible 
(e.g., M81, M87, M104), the AGN content in the majority of LINERs 
remains unknown.  Determining the physical origin of LINERs is more than of 
mere phenomenological interest.  Because LINERs are so numerous, they 
have a tremendous impact on the specification of the faint end of the local 
AGN luminosity function, which itself bears on a range of issues.
A number of recent developments provide considerable new insight into the 
origin of LINERs.  I outline these below, and I use them to advance the 
proposition that {\it most} LINERs are truly AGNs\footnote{Note that this 
paper is concerned only with compact, {\it nuclear} LINERs ($r$ \lax 200 pc), 
which are most relevant to the AGN issue.  LINER-like spectra are 
often also observed in extended nebulae such as those associated with 
cooling flows, nuclear outflows, and circumnuclear disks.}.

\subsection{Host Galaxy Properties}

LINERs and Seyferts live in virtually identical host galaxies (Fig. 2).
The vast majority of both classes occupy bulge-dominated, early-type 
systems (87\% are found in types E--Sbc), which clearly differ from the 
population of galaxies whose nuclear spectrum indicates photoionization by 
current star formation (the so-called H~II-nuclei), which is dominated by 
late-type hosts (63\% are Sc's and later).  The only noticeable difference 
in the distribution of morphological types of LINERs and Seyferts is that 
LINERs occupy a higher proportion of ellipticals.  Bars exist with roughly the
same frequency within the subsample of disk galaxies in both groups.

The similarity in the host galaxy properties of LINERs and Seyferts
becomes even more apparent when we examine their absolute
magnitude distributions (Fig. 2); they are statistically indistinguishable.
Both peak at $M_{B_T}^0\,\approx$ --20.5 mag (for $H_0$ = 75 \kms\ 
Mpc$^{-1}$), about 0.4 mag brighter than $M_{B_T}^*$, the typical absolute 
magnitude of the field-galaxy luminosity function.  The parent galaxies of 
H~II-nuclei, on the other hand, are systematically fainter than the other 
two groups by $\sim$0.5 mag in the median.

\subsection{Detection of Massive Black Holes}

There has been considerable recent progress in the detection of dark masses,
plausibly interpreted as massive black holes, in nearby galactic nuclei 
(see Ho 1998c and the contribution by S. Faber).  A significant fraction of the 
known black hole candidates, albeit still a small number, in fact are well 
known LINERs.  These include M81 ($M_{\rm BH}\,\approx$ 4\e{6} \solmass), 
M84 (1.5\e{9} \solmass), M87 (3\e{9} \solmass), the ``Sombrero'' galaxy 
(1\e{9} \solmass), NGC 4261 (5\e{8} \solmass), and Arp 102B (2\e{8} 
\solmass).  Although certainly no statistical conclusions can yet be drawn, 
these examples nevertheless serve as a powerful proof-of-concept that at 
least {\it some} LINERs are incontrovertibly accretion-powered sources.

\subsection{Detection of Broad-Line Regions}

\begin{figure}
\vbox{
\hbox{
\hskip 0.1truein
\psfig{file=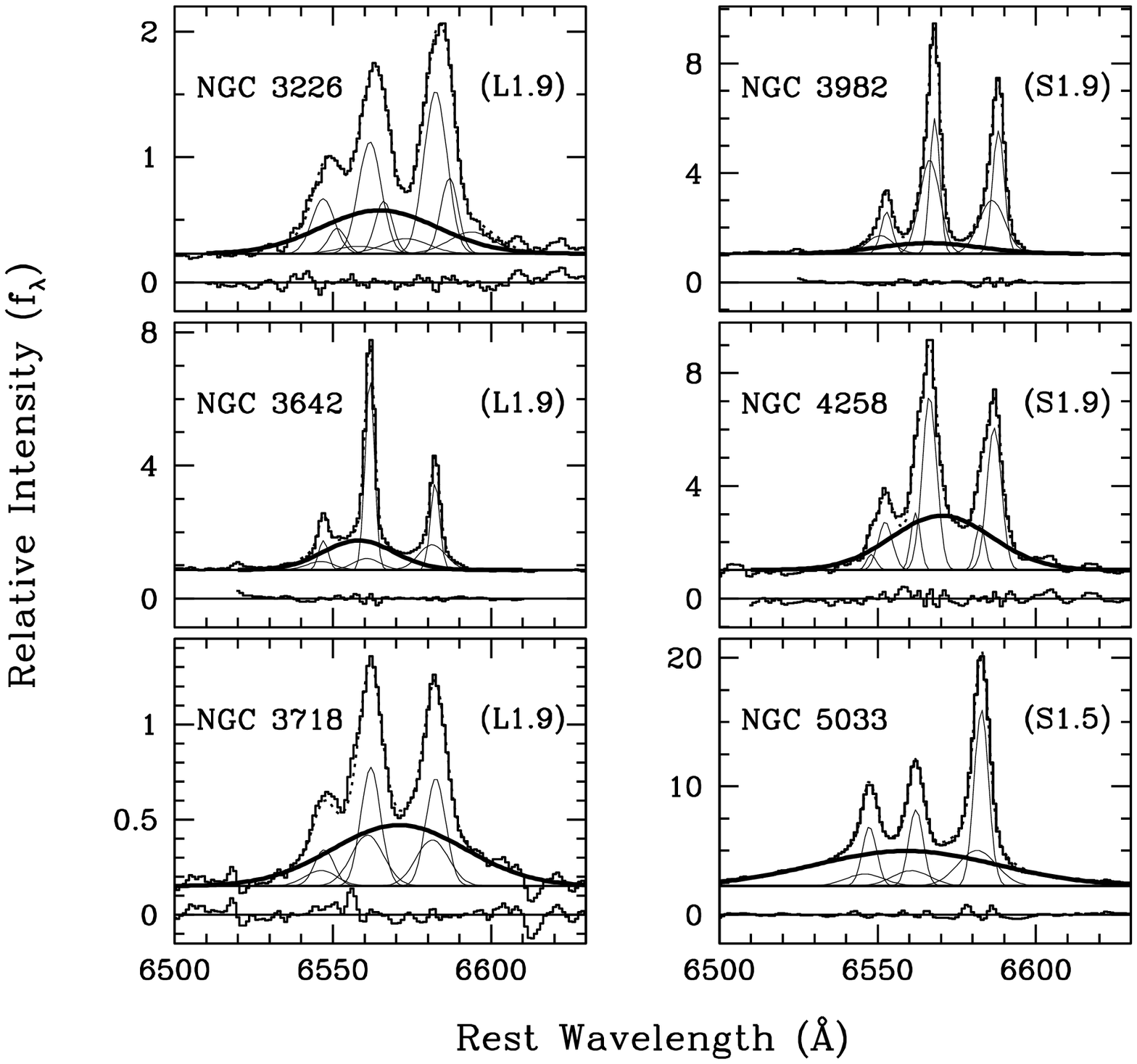,height=3.2truein,width=3.1truein,angle=0}
\hskip 0.4truein
\psfig{file=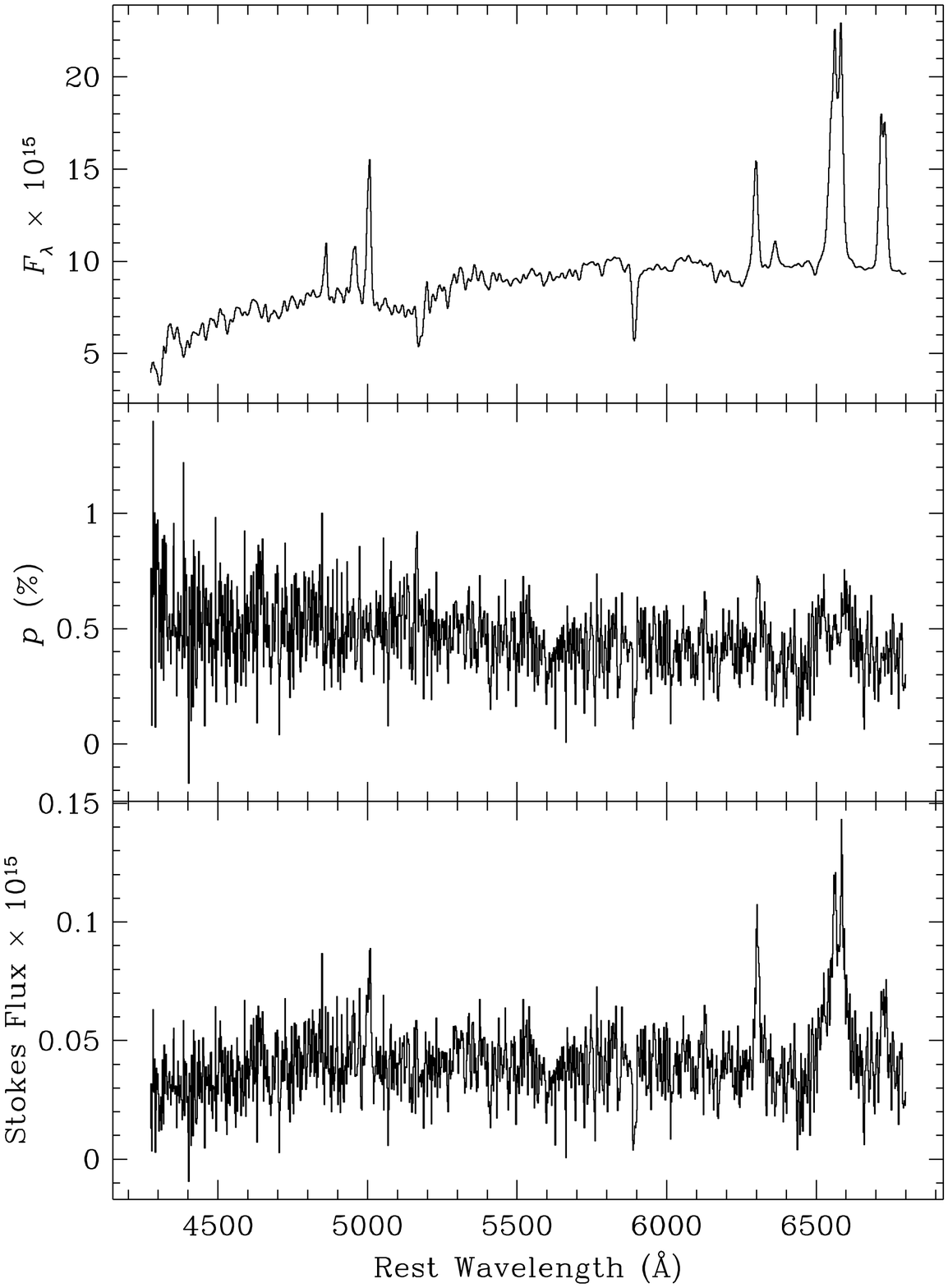,height=3.2truein,width=3.1truein,angle=0}
}
}
\vskip +0.5cm
Fig. 3 ({\it Left}).  Decomposition of the H\al\ + N~II region for LINERs 
and Seyferts.  The broad H\al\ component is shown as a 
heavy line.  Adapted from Ho \etal (1997b).
\vskip +0.1cm
Fig. 4 ({\it Right}).  Keck spectra of NGC 1052 from Barth (1998).  Top panel 
--- Total flux spectrum.  Middle panel --- Percent polarization.
Bottom panel --- Stokes flux obtained from $F_{\lambda} * p$.
\end{figure}

{\it Bona fide} AGNs such as QSOs and luminous Seyfert 1 nuclei distinguish
themselves unambiguously by their characteristic broad (FWHM $\sim$ few 
thousand \kms) permitted lines which arise from the broad-line region 
(BLR).  The detection of such broad lines in LINERs would constitute strong
evidence in favor of the AGN interpretation of these sources.  Since the 
strongest permitted line at optical wavelengths is expected to be H\al, 
one of the primary goals of the Palomar survey was to search for broad H\al\ 
emission.  Of the sample of objects with broad H\al\ emission (22\% of the AGN 
candidates), more than half belong to the LINER category (Ho \etal 1997b; see 
Fig. 3).  This is a very important finding, because it implies that LINERs, 
like Seyferts, evidently come in two flavors --- some have a visible BLR, and 
others do not.  By direct analogy with the nomenclature established for 
Seyferts, we might extend the ``type 1'' and ``type 2'' designations 
to include LINERs.  Approximately 15\%--25\% of the LINER population are LINER 
1s, the appropriate fraction depending on whether the so-called transition 
objects (Ho \etal 1993) are regarded as LINERs.

A remaining, outstanding question, however, is what fraction of the 
LINER 2s are AGNs.  Again, by analogy with the Seyfert 2 class, surely 
{\it some} LINER 2s must be genuine AGNs --- that is, LINERs that happen to 
have no BLR or have an obscured BLR.  There is no {\it a priori} reason why the 
unification model, which has enjoyed such popular support in the context of 
Seyfert galaxies, should not equally apply to LINERs.  The existence of an 
obscuring torus does not obviously depend on the value of the ionization 
level of the line-emitting regions.  If we suppose that the ratio of LINER 2s 
to LINER 1s is the same as the ratio of Seyfert 2s to Seyfert 1s, that ratio 
being 1.4:1 in the Palomar survey, we might argue that the AGN fraction in 
LINERs may be as high as $\sim$40\%--60\%.  
What evidence is there, however, that the unified model is applicable to 
LINERs?  The faintness of the sources in question renders application of the 
classical spectropolarimetric test (e.g., Antonucci and Miller 1985) 
impractical for moderate-sized telescopes.  An important breakthrough was 
recently achieved by Barth (1998), who successfully used the Keck 10~m 
telescope to detect a polarized broad H\al\ line in the prototypical LINER 
NGC 1052 (Fig. 4).  Weak broad H\al\ wings were previously found in the 
total-light spectrum after very careful profile decomposition (Ho \etal 
1997b), but the broad line is undeniable in scattered light.  Some LINER 
2s evidently {\it do} harbor obscured BLRs.  It would be highly desirable to
extend these observations to larger samples.


Lest one doubts that the existence of BLRs can be established with the 
detection of a single broad line, it should be remembered that broad lines 
are seen in other transitions as well, particularly in the ultraviolet (UV) 
where contamination by old stars poses less of a problem. The two best examples
are M81 (Ho \etal 1996) and NGC 4579 (Barth \etal 1996) which were observed 
with the {\it Hubble Space Telescope (HST)}.  Finally, note that 
the minority of AGNs that display so-called double-peaked broad emission 
lines, whose origin is widely thought to lie in a relativistically rotating 
accretion disk, in fact very often exhibit LINER-like narrow-line spectra 
(Eracleous 1998 an references therein).

\subsection{Ultraviolet Emission and Constraints on Shock Excitation}

The nonstellar nature of LINERs might be revealed through the presence of
a central compact source responsible for the photoionizing continuum.  The UV 
band is preferred over the optical because it minimizes contamination from 
old stars, although it is much more adversely affected by dust extinction.  
Two imaging surveys performed with the {\it HST} (Maoz \etal 1995; Barth \etal 
1998) find that LINERs in fact do contain compact UV emission, but in only 
20\%--25\% of the cases.  By itself, however, this result is ambiguous.
Are the central UV sources in most LINERs obscured by dust, are they in the 
``off'' state of a duty cycle most of the time as suggested by Eracleous \etal
(1995), or do the majority of LINERs simply lack a 
pointlike ionizing source because they are not AGNs after all?   There is some 
indication that the sources detected in the UV tend to be in more face-on
galaxies than the undetected sources (Barth \etal 1998).  Moreover, as 
discussed below, LINERs seem to be {\it intrinsically} weak in the UV, and this
may further contribute to the low detection rates.  Mere 
morphological information, of course, cannot specify definitively the 
physical origin of the UV emission.  For example, the point sources could be 
simply very compact nuclear star clusters.  Indeed, follow-up spectroscopy 
indicates that the bulk of the UV emission in some sources comes from 
young massive stars (Maoz \etal 1998).  Others, on the other hand, 
exhibit featureless, power-law continua as expected for an energetically 
significant AGN component (M81: Ho \etal 1996; NGC 4579: Barth \etal 1996; 
M87: Tsvetanov \etal 1998).

Collisional ionization by shocks has been considered a plausible energy 
source for LINERs since the discovery of these objects (Fosbury \etal 1978; 
Heckman 1980).  Dopita and Sutherland (1995) recently showed that the diffuse 
radiation field generated by fast ($v\,\approx$ 150--500 \kms) shocks can 
reproduce the optical narrow emission lines seen in both LINERs and Seyferts.  
In their models, LINER-like spectra are realized under conditions in which the 
precursor H~II region of the shock is absent, as might be the case in 
gas-poor environments.  The postshock cooling zone attains a much higher 
equilibrium electron temperature than a photoionized plasma; consequently, a 
robust prediction of the shock model is that it should produce a 
higher excitation spectrum, most readily discernible in the UV, than 
photoionization models.  In all the cases studied so far, the UV spectra 
are inconsistent with the fast-shock scenario because the observed intensities 
of the high-excitation lines such as C~IV \lamb1549 and He~II \lamb1640 are 
much weaker than predicted (Barth \etal 1996, 1997; Nicholson \etal 1998; 
Maoz \etal 1998).  [The case of M87 presented by Dopita \etal (1997) is 
irrelevant to the present discussion because those observations explicitly 
avoided the nucleus of the galaxy.]  The data, however, cannot rule out 
contributions from slower shocks ($v$ \lax 150 \kms), although the viability 
of shock ionization in luminous AGNs has been criticized on energetic grounds 
by Laor (1998).

\subsection{Clues from the X-rays}

Compact soft X-ray emission on the scale of the {\it ROSAT} HRI camera 
($\sim$5\asec) has now been detected in a handful of LINERs (e.g., Worral 
and Birkinshaw 1994; Koratkar \etal 1995; Fabbiano and Juda 1997), although 
no statistical conclusions can yet be drawn based on the scant data available.  
Most of the core sources have luminosities clustering near $L$(0.5--2 keV) 
$\approx$ 10$^{40}$--10$^{41}$ \lum\ because of selection effects.  The 
pointlike morphology of the {\it ROSAT} images certainly agrees with our 
expectation for an AGN source, but we must remember that the 5\asec\ 
point-spread function of the HRI subtends an uncomfortably large region 
(several hundred parsecs) at the typical distances of these objects.  Images
taken at much higher angular resolution and ideally at harder energies, 
such as would be possible with the ACIS camera on {\it AXAF} (see the 
concluding remarks at the end of the paper), are needed to put more stringent 
constraints on the nature of the X-ray emission.

In the meantime, progress can be made by examining the {\it ASCA} hard X-ray 
spectra of LINERs whose X-ray structure is found to be compact on HRI images.
These data, again, are scarce, and current constraints by necessity bias the 
sample in favor of the brightest targets.  Nonetheless, when the observations 
are considered collectively (Serlemitsos \etal 1996; Terashima \etal 1997; 
Ptak 1998; Ptak \etal 1998; Iyomoto \etal 1998; Nicholson \etal 1998;
see contributions by H. Awaki and Y. Terashima), the following trends appear.  
(1) The 2--10 keV 
continuum can generally be modeled as a single power-law function modified by 
cold absorption; the best-fitting photon index, $\Gamma\,\approx$ 1.7--1.8, 
agrees well with values normally measured in luminous AGNs.  In some cases 
the fits require an additional soft thermal component with a temperature of 
$kT\,\approx$ 1 keV.  (2) There is no evidence for significant amounts of 
cold material along the line of sight.  Any measurable absorbing column in 
excess of the Galactic contribution usually does not exceed 
$N_{\rm H}\,\approx$ 10$^{21}$ \percm2.  (3) Broad Fe K\al\ emission at 6.4 
keV, a feature common to many luminous Seyfert 1 nuclei, is usually either 
absent or unusually weak.  The composite LINER spectrum of Terashima \etal 
(1997) shows no detectable Fe K\al\ line to an equivalent-width limit of 140 
eV. In the few cases where an iron line has been detected, the rest energy 
is $\sim$6.7 keV, consistent with ionized instead of neutral iron.  And (4), 
these sources do not undergo rapid X-ray variability at the level expected 
from extrapolation of the variability behavior established for more luminous 
sources.


\subsection{Compact Radio Cores}

Finally, further verisimilitude with the AGN phenomenon can be sought 
by means of high-resolution radio continuum observations.  As already 
mentioned above, compact, flat-spectrum cores of low power are 
often found in radio continuum surveys of nearby elliptical and S0 galaxies.
Less certain is the incidence of the similar phenomenon in spirals, which 
in general tend to be rather weak radio sources.  A large subsample of 
LINERs from the Palomar survey is being systematically mapped at 2, 3.6, 
and 6~cm using the the Very Large Array in its most extended configuration.
The highest angular resolution achieved is comparable to that of {\it HST}, 
and faint, sub-mJy compact sources can be routinely detected with ease.  
The preliminary analysis, reported in Van Dyk and Ho (1997), finds that the
6~cm maps nearly always detect a single, compact core spatially coincident 
with the optical nucleus.  Most of the radio cores have relatively steep 
spectral indices consistent with optically thin synchrotron emission, but a 
significant fraction has flat, presumably optically thick, spectra 
(Falcke \etal 1997).  

\subsection{The Spectral Energy Distributions of LINERs}

\begin{figure}
\hskip 2.3truein
\psfig{file=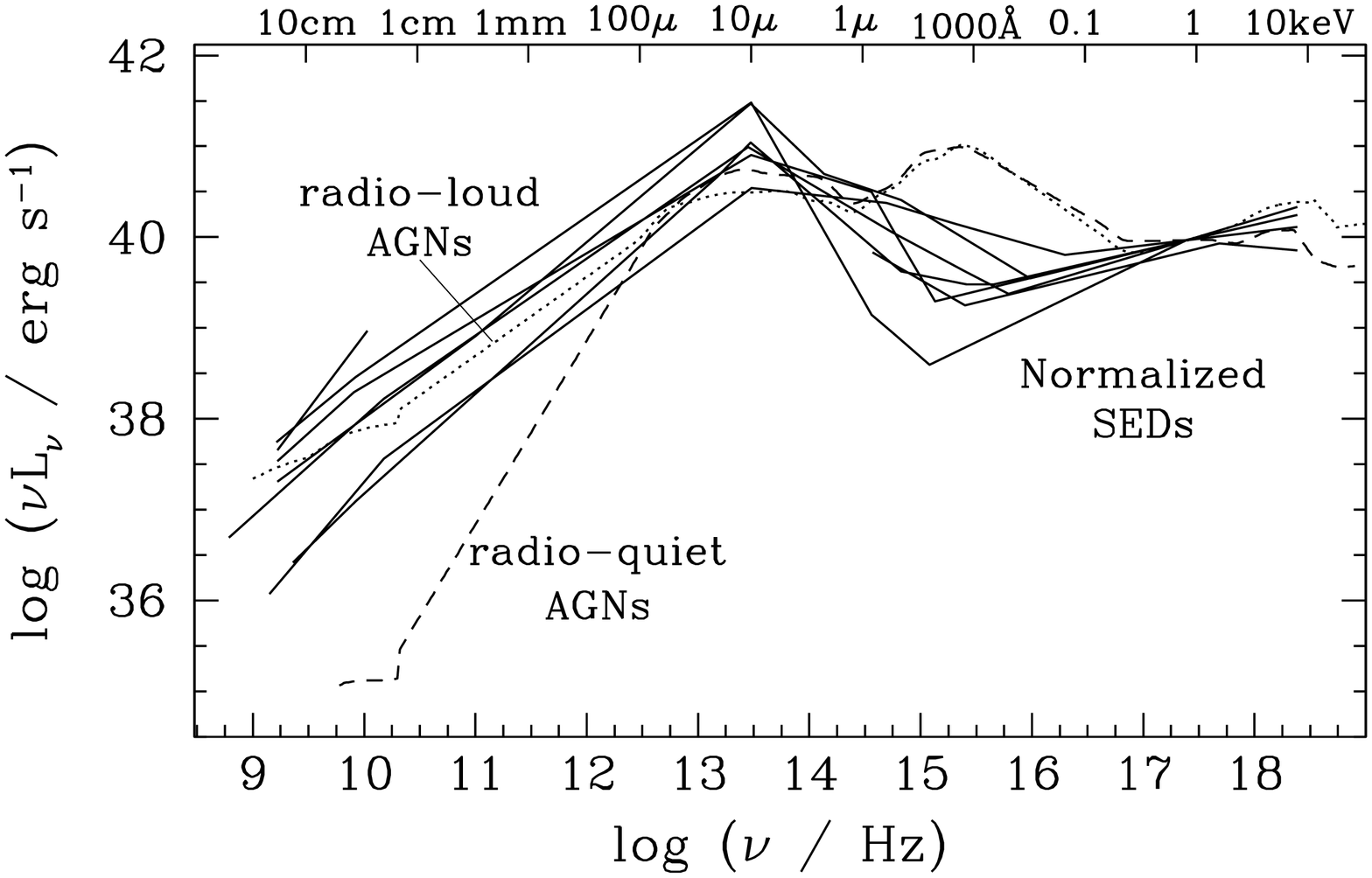,height=3.0truein,angle=270}
\vskip +0.5cm
Fig. 5.  Interpolated SEDs of seven low-luminosity AGNs ({\it solid lines})
normalized to the 1 keV luminosity of M81.  The median SEDs of radio-loud
({\it dotted line}) and radio-quiet ({\it dashed line}) AGNs of Elvis \etal
(1994), normalized in the same way, have been overplotted for comparison.
From Ho (1998b).
\end{figure}

Luminous AGNs generally display a fairly ``universal'' spectral energy 
distribution (SED) (e.g., Elvis \etal 1994).  The SED from the infrared to the 
X-rays, roughly flat in log~$\nu F_{\nu}$--log~$\nu$ space, can be represented 
by an underlying power law ($\alpha\,\approx$ 1, where 
$F_{\nu}\,\propto\,\nu^{-\alpha}$) superposed with several distinct 
components, the most prominent of which is a broad UV excess.  This so-called 
big blue bump is conventionally interpreted as thermal emission from an 
optically thick, physically thin accretion disk (Malkan and Sargent 1982).  
As multiwavelength data for LINERs and other low-luminosity AGNs become
more readily available, we can begin to piece together their SEDs.  Comparing
SEDs of AGNs of various luminosities might yield clues to physical processes 
that depend on luminosity.

The SEDs of the low-luminosity AGNs share a number of common traits, and yet 
they differ markedly from the SEDs of luminous AGNs (Ho 1998b).  To 
illustrate this point, Figure 5 normalizes the SEDs of seven low-luminosity 
objects (mostly LINERs) with the median SED of radio-loud and radio-quiet AGNs 
from Elvis \etal (1994).  Several features are noteworthy.  (1)  The 
optical-UV slope is quite steep.  The power-law indices for the seven 
low-luminosity objects average $<\alpha>\,\approx\,1.8$, whereas in luminous
AGNs $\alpha \,\approx$ 0.5--1.  (2) The UV band is exceptionally dim relative 
to the optical and X-ray bands; there is no evidence for a big blue bump 
component.  Indeed, the SED reaches a local minimum somewhere in the far-UV or 
extreme-UV region.  The mean value of $\alpha_{\rm ox}$, the two-point 
spectral index between 2500 \AA\ and 2 keV, is $\sim$0.9, to be compared with
$<\alpha_{\rm ox}>$ = 1.2--1.4 for luminous Seyferts and QSOs.  (3) There is 
tentative evidence for a maximum in the SED at mid-IR wavelengths.  
(4) The nuclei have radio spectra that are either flat or inverted.  
(5) All sources, including the three spiral galaxies in the sample, qualify as 
being {\it radio-loud}.  This finding runs counter to the usual notion that 
only elliptical galaxies host radio-loud AGNs.  (6) The bolometric 
luminosities of the sources range from $L_{\rm bol}$  = 2\e{41} to 8\e{42} 
\lum, or $\sim$$10^{-6}-10^{-3}$ times the Eddington rate for the black hole 
masses estimated for these objects.

The overall characteristics of these nonstandard SEDs can be explained 
by models of ``advection-dominated accretion flows'' (ADAFs; 
see Narayan \etal 1998 for a review).  The accretion flow equations admit 
a stable advection-dominated, optically thin solution when the accretion 
rate falls to very low values ($\dot M$ \lax $10^{-2} \dot M_{\rm Edd}$).  
Under these conditions, the low density 
and low optical depth of the accreting material lead to inefficient 
cooling, and the resulting radiative efficiency is much less than the 
canonical value of 10\%.  This accounts for the low observed luminosities. 
Moreover, the predicted SEDs of ADAFs look radically different from the 
SEDs of optically thick disks but provide a good match for the observations 
of low-luminosity AGNs (Ho and Narayan 1998).

\section{A PHYSICAL DESCRIPTION OF LINERs}

The evidence summarized in the preceding sections paints the 
following picture for the physical nature of LINERs.  For reasons yet to 
be fully understood, the narrow emission-line regions of AGNs evidently can 
allow a wide range of ionization parameters (the ratio of the density of 
ionizing photons to the density of gas at the illuminated face of a cloud; 
$U\,\equiv\,L_{\rm ion}/4 \pi c n r^2$).  Those with $U\,\approx$ 
10$^{-2}$--10$^{-1}$ are conventionally denoted ``Seyferts;'' low-ionization 
objects  with $U\,\approx$ 10$^{-3.5}$--10$^{-2.5}$ are called ``LINERs'' 
(Halpern and Steiner 1983; Ferland and Netzer 1983; Ho \etal 1993).  
There is no sharp division, other than that imposed for sheer taxonomical 
convenience, between the two groups.  Like Seyferts, some LINERs ($\sim$20\%) 
come equipped with a BLR.  The others either do not have BLRs, or their BLRs 
are obscured from the observer.  At least some of the type 2 LINERs definitely 
have hidden BLRs that can only be seen in scattered light.

Most LINERs inhabit large, bulge-dominated galaxies, the very galaxies that 
evidently are most prone to host supermassive black holes.  Thus, LINERs can be 
identified with the quiescent black-hole remnants from the quasar era.  
In the present epoch, the supply of gas available for powering the 
central engines is much curtailed, and the resulting low accretion rates 
lead to an advection-dominated mode of accretion that manifests itself 
in the low luminosity output and in the peculiar SEDs of the nuclei.  The 
diminished nuclear power naturally accounts for the low values of $U$ in 
LINERs, and the likely preponderance of ADAFs in nearby galactic nuclei 
may explain why LINERs are so ubiquitous.  Two additional effects may help to 
further reduce the ionization 
state of the line-emitting gas.  The modification of the SED from UV to 
X-ray energies, most noticeable by the absence of the UV excess, leads to an 
overall hardening of the ionizing spectrum.  The exceptional strength of the 
radio component, on the other hand, increases the effectiveness of cosmic-ray 
heating of the gas by the energetic electrons (e.g., Ferland and Mushotzky 
1984).  Both effects result in an enhancement of the low-ionization lines.

\section{FUTURE DIRECTIONS}

Future studies should refine the statistical completeness of the current AGN 
surveys.  The Palomar survey, while a significant improvement compared to 
previous optical surveys, is nonetheless limited by ground-based seeing and 
by host galaxy contamination.  In early-type galaxies, where the contamination 
from starlight is strong, there is a practical limit to which weak emission 
lines can be extracted from the total spectrum.  On the other hand, in
late-type galaxies the strong emission from H~II regions in and near the 
nucleus can easily mask the fainter signature of a weak AGN that might be 
present.  The reported low incidence of AGNs in late-type galaxies, therefore,
may be misleading and needs to be verified.  A significant increase in 
sensitivity to weak nuclear emission can be achieved with high-angular 
resolution spectroscopy with {\it HST}.  A follow-up program to assess the 
completeness of the Palomar survey is being performed with the Space Telescope 
Imaging Spectrograph.

The recent availability of high-quality multiwavelength observations has 
provided important clues toward resolving the LINER mystery.  As outlined in 
this review, while the data collectively, and in some instances even 
individually, do support the AGN interpretation for LINERs, it is still 
premature to draw quantitative, statistical conclusions concerning the AGN 
content in LINERs.  The most outstanding unanswered question is the fraction 
of the LINER 2 population which are genuine AGNs.  The cleanest test of the 
AGN hypothesis must rely on high-resolution observations in the hard X-ray 
band, which is the least affected by absorption and by confusion from young 
stars.  Detection of a single compact hard X-ray source coincident with the 
nucleus would constitute strong evidence for the existence of an 
accretion-powered source.  Such an experiment is being planned for the 
ACIS camera (resolution $\sim$0\farcs5) on {\it AXAF}.

\acknow

My work is supported by NASA grants from the Space Telescope
Science Institute, which is operated by AURA, Inc., under NASA contract 
NAS5-26555.  This paper was written during a visit to the Academia Sinica's 
Institute of Astronomy and Astrophysics; I thank the members of the Institute, 
and especially its director, K.~Y. Lo, for their warm hospitality.

\end{document}